\documentclass[12pt]{article}
\newif\iffigs\figstrue
\usepackage{latexsym}
\iffigs
  \input{epsf}
\else
\fi
 \font\smallmsbm=msbm10 scaled 800
 \font\bigmsbm=msbm10 scaled 1800
 \font\tenmsbm=msbm10 scaled 1200 
 \font\sevenmsbm=msbm9 
 \newfam\msbmfam
 \textfont\msbmfam=\tenmsbm
 \scriptfont\msbmfam=\sevenmsbm
 
 \def\msbm{\fam\msbmfam\tenmsbm}
 \def\msbmn{\fam\msbmfam\sevenmsbm}
 \def\msbms{\fam\msbmfam\smallmsbm}
\makeatletter
\@addtoreset{equation}{section}
\makeatother
\renewcommand{\theequation}{\thesection.\arabic{equation}}
\newcounter{parentequation}
\newenvironment{subequations}{%
  \refstepcounter{equation}%
  \begingroup 
  \let\protect\noexpand
  \edef\@tempa{\def\noexpand\theparentequation{\theequation}}%
  \expandafter
  \endgroup\@tempa
  \setcounter{parentequation}{\value{equation}}%
  \setcounter{equation}{0}%
  \def\theequation{\theparentequation\alph{equation}}%
  \ignorespaces
}{%
  \setcounter{equation}{\value{parentequation}}%
}

\newcommand{\eqn}[1]{(\ref{#1})}
\newsavebox{\uuunit}
\sbox{\uuunit}
    {\setlength{\unitlength}{0.825em}
     \begin{picture}(0.6,0.7)
        \thinlines
        \put(0,0){\line(1,0){0.5}}
        \put(0.15,0){\line(0,1){0.7}}
        \put(0.35,0){\line(0,1){0.8}}
       \multiput(0.3,0.8)(-0.04,-0.02){10}{\rule{0.5pt}{0.5pt}}
     \end {picture}}

\def\IC{\hbox{\msbm C}}
\def\bfzero{\relax\,\hbox{$\inbar\kern-.3em{\rm 0}$}}

\def\IZ{\hbox{\msbm Z}}
\def\IP{\hbox{\msbm P}}
\def\sIP{\hbox{\msbmn P}}
\def\bfone{\relax{\rm 1\kern-.35em 1}}
\def\cA{{\cal A}} 
 
 \def\cS{{\cal S}}

\def\cL{{\cal L}} 
\def\cN{{\cal N}} \def\cO{{\cal O}}

\newcommand{\dfrac}{\displaystyle \frac}
\def\beq{\begin{equation}}
\def\eeq{\end{equation}}
\def\bea{\begin{eqnarray}}
\def\eea{\end{eqnarray}}
\def\bet{\begin{tabular}}
\def\eet{\end{tabular}}
\def\bes{\begin{subequations}\begin{eqnarray}}
\def\ees{\end{eqnarray}\end{subequations}}

\def\a{\alpha}

\begin{document}
\begin{titlepage}

\begin{flushright}
DFTT 99/23 \\
hep-th/9904198\\
April 1999\\
\end{flushright}

\vspace{2truecm}

\begin{center}

{ \Large \bf ${\cal N} = 2$ conformal field theories from $M2$--branes 
at conifold singularities.}

\vspace{1cm}

{Gianguido Dall'Agata$^{\dag}$}\footnote{dallagat@to.infn.it}  
\vspace{1cm}%

{$\dag$ \it Dipartimento di Fisica Teorica, Universit\`a degli 
studi di Torino \\
and \\
Istituto Nazionale di Fisica Nucleare, Sezione di Torino, \\
 via P. Giuria 1, I-10125 Torino.}

\medskip


\vspace{1cm}

\begin{abstract}
We make some comments on the derivation of $\cN = 2$ super--conformal 
field theories with smooth gauge group from 
$M2$--branes placed at conifold singularities, giving a detailed 
prescription for two specific examples: the singular cones over the $Q^{111}$ 
and $M^{110}$ manifolds.

\end{abstract}

\end{center}
\vskip 4.5truecm 
\end{titlepage}
 
\newpage

\baselineskip 6 mm


\section{Introduction}

The most important advances in the knowledge of the non--perturbative 
aspects in string--theory have been due to the discovery and the study 
of $D$--branes and $M$--theory. 
The first have revealed to be essential in the construction of a 
consistent theory of strings and also very important to test geometry 
at sub--stringy scale. 
The latter has opened the possibility to have interesting informations 
on the strong coupling limit of such theories.

A striking example of the new aspects discovered by considering 
$D$--branes as probes of stringy interactions is that 
space--time is a derived rather than a 
primary concept, arising as the moduli space of $D$--brane 
world--volume gauge theories.
Some specific analyses \cite{DGM,Quivers,inDGM}, where 
branes are placed at orbifold singularities, show indeed such 
an unexpected feature.

Another important developement that has recently emerged is the possible 
relation between certain gauge theories, describing world--volume 
theories of large numbers of $D$ or $M$--branes, and 
superstring or $M$--theory on backgrounds of the factorised form 
$AdS \times H$, with $H$ a compact manifold \cite{Maldacena, 
GubserKlebanovPolyakov}.

In flat space the conjectured equivalence relation specifies $H$ to 
be a sphere, as $AdS_{p+2} \times S^{d-p-2}$ arises as the {\it 
horizon} manifold for a $Dp$ or $Mp$--brane in $d$--dimensions.
This is no longer true if we place the branes on singular spaces.
Since there is no obstruction to placing the branes at the singular 
point, the horizon $H$ will now be different from the one we had at 
smooth points (i.e. the sphere).
This leads to the possibility of deriving superconformal gauge 
theories with less than the maximal supersymmetry.

In string theory, the SCFT describing $D$--branes at 
orbifold singularities can be derived by projecting out the invariant 
states and the potential from the flat space theory \footnote{
In \cite{Fre1}it was considered for the first time the orbifold limit of ALE 
manifolds (or Taub Nut, or higher dimensional generalizations) as a 
device to define the corresponding brane conformal field theory.
} 
\cite{DGM,Quivers, Greene}.
On other kind of singularities, such as 
conifolds, one must find others and indirect ways of derivation. 
Among many attempts regarding this kind of singularities 
\cite{KW, altri},  the most satisfying way seems to be the one 
outlined in \cite{MorrisonPlesser}.
If one can find an orbifold, from which a partial resolution leads to 
the desired singularity, the corresponding gauge theory can be derived from the 
orbifold one by giving appropriate expectation values to some of the 
moduli fields and studying the reduced theory.
The choice is obviously related to the way of resolving the orbifold 
singularity and it is controlled by the Fayet--Ilioupulos terms 
(governed by the twisted string sectors).

For $D3$--branes in type $IIB$ string theory, the relevant 
singularities are the Gorenstein canonical singularities in ${ 
\IC}^3$ and a wide class is given by those which can be described by 
toric geometry\footnote{All singularities of the form ${\hbox{\msbmn 
C}}^n
/\Gamma$ for finite abelian groups $\Gamma \subset U(n)$ are toric 
singularities.} some of which have been carefully studied in 
\cite{MorrisonPlesser}.

A natural generalization would be to study the fundamental branes of 
$M$--theory at $\IC^4/\Gamma$ singularities. 
This could lead to three--dimensional theories 
with ${\cN} \leq 8$.

There are many problems connected with this programme, from the 
absence of a complete classification of such Gorenstein singularities, 
to the possibility of having singular horizons \cite{inMP}.
Good and interesting cases are the ones where the horizon is given by 
a coset space of dimension seven.
Generic coset manifolds admitting an Einstein metric have been described 
in \cite{Jensen}, while the seven--dimensional ones have been 
completely classified in \cite{CRW}. 
These spaces can be retrieved as horizon 
manifolds for $M2$--branes placed at the singular point of 
a space of the form $M_3 \times C(H)$ where $C(H)$ is 
the eight--dimensional cone over the coset--space \cite{Duffetal}.

In this paper we deal with such singular cones when the corresponding 
supergravity theory preserves $\cN = 2$ supersymmetry.
The language we will use is that of toric geometry.
This is indeed an effective way of describing the classical $D$--brane 
moduli space as pointed out in \cite{DGM, Infirri} and also a very 
simple tool to deal with ADE singularities.
For an introduction to toric geometry for physicists see \cite{Intro}. 
A more rigorous exposition can be found in the book by Fulton \cite{Fulton} and 
recent developements are exposed in \cite{Cox}.
Some useful considerations about equations defining toric varieties 
and combinatorial aspects can be found  in 
\cite{Sturm1,Sturm2}.

In the next section we are going to describe the $\cN = 2$ $C(H)$ cones 
in this language, specifying the charge matrices for their 
definition as a symplectic quotient.
In section three we show which are the classes of Gorenstein ${ 
\IC}^4/ \Gamma$ singularities allowed by the physical consistency 
requirements and show which resolutions can lead to the conifold 
singularities we want to describe.
The last section contains some comments on the related $\cN = 2$ 
field theories.

\section{The cones over coset manifolds}

Eleven--dimensional supergravity can be 
spontaneously compactified through the Freund--Rubin mechanism to a 
space of the form $AdS_4 \times H^7$, where $H^7$ is one of the coset 
spaces classified in \cite{CRW}.
The cases we want to analyse are the $\cN = 2$ ones, namely the 
$Q^{ppp}$ \cite{N1}, the $M^{ppr}$ \cite{N2} and the $V_{5,2}$,
which are defined as the following group quotients
$$
\frac{SU(2) \times SU(2) \times SU(2)}{U(1) \times U(1)}, 
\quad
\frac{SU(3) \times SU(2) \times U(1)}{SU(2) \times U(1)},
\quad \hbox{ and } \quad 
\frac{SO(5)}{SO(3)}.
$$
For the sake of simplicity, for the first two classes of manifolds, 
we are going to choose a single 
representative given by $Q^{111}$ and $M^{110}$.
It is worth pointing out that all the other $M^{ppr}$ manifolds can 
be derived from $M^{110}$ by quotienting with a suitable cyclic group 
\cite{CRW}.

Since we want to use the language of toric geometry we have to give a 
description of these manifolds in such a language.
The starting point to make this construction is a theorem reported in 
\cite{BoyGal} which states that {\it if $\cS$ is a compact quasi-regular
homogeneous Sasakian--Einstein manifold, then $\cS$ is a circle-bundle over a
generalized flag manifold.}
Since our seven--dimensional manifolds are Sasaki--Einstein we can 
apply such a theorem.
The $Q^{111}$ manifold could be recovered as a circle bundle over 
$\IP^1 \times \IP^1 \times \IP^1$, $M^{110}$ fibering over $\IP^2 
\times \IP^1$ and the Stiefel manifold $V_{5,2}$ fibering over the 
real Grassmanian of projective lines in the real $\IP^4$.

The essential requirement is that these manifolds {\it admit} a toric description.
We will see that $Q^{111}$, $M^{110}$ and their cones can be easily described 
by toric geometry, but
this is not the case for the Stiefel manifold. 
This manifold can be described \cite{KW} as the following surface 
embedded in $\IP^4$:
\beq
\label{quadric}
\sum_{i=1}^5 {z_i^2} = 0.
\eeq
We also know that a toric manifold can always be described by 
embedding equations given by one monomial equals another monomial.
Since the quadric \eqn{quadric} cannot have such a description in 
$\IP^4$ without a change in its degree, 
we conclude that this is not a toric variety.

We thus limit ourselves to the $Q$ and $M$ cases.


\subsection{Toric description of $Q^{111}$}

The first manifold we describe is the $Q^{111}$, which can be found 
as the unit circle bundle inside the $\cL_Q = \cO_{\sIP^1} (-1) \otimes 
\cO_{\sIP^1} (-1) \otimes \cO_{\sIP^1} (-1)$ line bundle.

$(\hbox{\msbm P}^1)^3$ as a toric variety can be described by a fan 
generated by $\{\pm e_1, \pm e_2, \pm e_3\}$.
From the fan, one can simply deduce the matrix 
describing the combinatorics for this variety
$$
 \left( \bet{cccccc} 
1 & 1 & 0 & 0 & 0 & 0 \\
0 & 0 & 1 & 1 & 0 & 0 \\
0 & 0 & 0 & 0 & 1 & 1 
\eet
\right)
$$
and one can also associate six homogeneous coordinates to this space as 
described in \cite{Coxcoord}.
In this case they are just the 
couples of homogeneous coordinates describing the three projective 
lines.

Line bundles over this space can be described in toric geometry by 
including a seventh {\it affine} coordinate.
The $\cL$ line--bundle is indeed described by the matrix
\beq
{\cal A} = \left( \bet{cccccc|c} 
$A_1$ & $A_2$ & $B_1$ & $B_2$ & $C_1$ & $C_2$ & $p$ \\\hline
1 & 1 & 0 & 0 & 0 & 0 & -1\\
0 & 0 & 1 & 1 & 0 & 0 & -1\\
0 & 0 & 0 & 0 & 1 & 1 & -1
\eet
\right), 
\eeq
which, with a change of basis, can be presented as 
\beq
\label{Amatrix}
{\cal A}^\prime = \left( \bet{cccccc|c} 
1 & 1 & 1 & 1 & -2 & -2 & 0\\
1 & 1 & -1 & -1 & 0 & 0 & 0\\
0 & 0 & 0 & 0 & 1 & 1 & -1
\eet
\right). 
\eeq

Since $Q^{111}$ is the circle bundle inside this $\IC^*$--bundle, 
it can be retrieved fixing the absolute value of $p$,
e.g. $|p|^2 = 1$.

Thus, $Q^{111}$ is the submanifold of the $\cal A$--generated manifold 
described by the $D$--term equations 
\beq
\label{hor}
|A_1|^2 + |A_2|^2 = |B_1|^2 + |B_2|^2 = |C_1|^2 + |C_2|^2 = 1,
\eeq
and quotiented by the two $U(1)$ actions given by the first two rows 
of \eqn{Amatrix}
\bes
\label{U1a}
(A,B,C) & \to & (e^{i \a} A, e^{i\a} B, e^{-2 i \a} C), \\
(A,B) & \to& (e^{i \a} A, e^{-i\a} B). 
\ees
It has therefore a description in terms of $(S^3 \times S^3 \times 
S^3)/(U(1) \times U(1))$.

At this point we can also build the invariant coordinates to obtain an explicit 
embedding of $C(Q^{111})$ inside $\IC^8$.
These coordinates, invariant under the $\IC^{*}$ action described by 
\eqn{Amatrix}, are
\beq
\begin{array}{rclrcl}
    z_0 &=& A_1 B_1 C_1; & z_4 &=& A_2 B_1 C_1;\\
    z_1 &=& A_1 B_1 C_2; & z_5 &=& A_2 B_1 C_2;\\
    z_2 &=& A_1 B_2 C_1; & z_6 &=& A_2 B_2 C_1;\\
    z_3 &=& A_1 B_2 C_2; & z_7 &=& A_2 B_2 C_2;\\
\end{array}
\eeq
and a set of independent embedding equations is thus given by
\beq
\left\{ 
\bet{rcl}
$z_0 z_5$ &=& $z_1 z_4,$ \\
$z_2 z_7$ &=& $z_3 z_6,$ \\
$z_0 z_3$ &=& $z_1 z_2,$ \\
$z_4 z_7$ &=& $z_5 z_6.$ 
\eet
\right.
\eeq

We would like to point out here that these can be viewed as the 
equations defining $Q^{111}$ in $\IP^7$ and that they are simply the 
equations for the Segre embedding  $(\IP^1)^3 \hookrightarrow \IP^7$.
$C(Q^{111})$ is then the affine cone over the 
projectively embedded variety.

In this description, $R$--symmetry corresponds to $p$--coordinate 
rotations.
Since we can use the last row $U(1)$ action to gauge--fix $p=1$, 
choosing an appropriate gauge for the remaining $U(1)$'s, we can 
deduce the $R$--symmetry charges of the various coordinate fields.
This shows that not all the coordinates have the same charges and 
therefore some of them must be composite states of the fundamental 
fields.

\subsection{Toric description of $M^{110}$}

As we have just done for the $Q^{111}$ manifold, we can give an 
analogous description of the $M^{110}$ manifold as 
the circle bundle inside the canonical line bundle
$\cL_M = \cO_{\sIP^2} (-3) \otimes \cO_{\sIP^1} (-2) \to \IP^{2} \times \IP^1$.

The fact that this is the correct line bundle can be easily derived by
the construction presented 
in \cite{CRW}, where a generic $M^{pq0}$ is seen as a $S^5
\times S^3 / U(1)$ manifold.
Indeed $S^5$ is a $U(1)$--bundle over $\IP^2$ and $S^3$ is a $U(1)$--bundle 
over $\IP^1$. 
The $U(1)$ at the denominator identifies the two bundles
 with an action of the type $(e^{2 i q \a} \, U, e^{-3 i p \a} V)$, where 
$U$ are the $\IP^2$ homogeneous coordinates and $V$ the $\IP^1$ ones.
It is straightforward then to derive the toric description of such a bundle
for $p=q=1$.

Repeating the construction of the last section,
$\IP^2 \times \IP^1$ as a toric variety can be described by a fan 
generated by $\{e_1, e_2, -(e_1+e_2), \pm e_3\}$.
From this, one can deduce the matrix encoding the combinatorics for 
this variety and describe the line--bundle $\cL$ as
\beq
{\cal A} = \left( \bet{ccccc|c} 
$U_1$ & $U_2$ & $U_3$ & $V_1$ & $V_2$ & $p$ \\\hline
1 & 1 & 1  & 0 & 0 & -3\\
0 & 0 & 0  & 1 & 1 & -2
\eet
\right).
\eeq

Again, with a change of basis we get
\beq
\label{AmatrixM}
{\cal A}^\prime = \left( \bet{ccccc|c} 
2 & 2 & 2 & -3 & -3 & 0\\
0 & 0 & 0  & 1 & 1 & -1
\eet
\right). 
\eeq

The $M^{110}$ manifold is then the horizon described by
\beq
\label{horM}
|U_1|^2 + |U_2|^2 + |U_3|^2 = 3, \qquad |V_1|^2 + |V_2|^2 = 2,
\eeq
and quotiented by the $U(1)$--action
\beq
\label{U1}
(U,V) \to (e^{2 i \a} U, e^{-3 i \a} V),
\eeq
exactly the desired $(S^5 \times S^3)/U(1)$ description,
 with the \eqn{U1} $U(1)$--action.  

With the same procedure of the last section, we 
can again build the invariant coordinates
\beq
\begin{array}{rclrcl}
    z_0 &=& U_1^3 V_1^2; &\ldots &&  \ldots\\
    z_1 &=& U_1^2 U_2 V_1^2; & z_{27} &=& U_1 U_2 U_3 V_2^2;\\
    z_2 &=& U_1 U_2 U_3 V_1^2; & z_{28} &=& U_2 U_3^2 V_2^2; \\
    \ldots & & \ldots & z_{29} &=& U_3^3 V_2^2;
\end{array}
\eeq
and the embedding equations in $\IC^{30}$, which we will not specify 
here\footnote{The embedding equations for the ${\cal L}_M$ line bundle have
 been independently derived by \cite{Fre2} who also claim that they 
describe the cone over $M^{11r}$ for any $r$.}.

\section{$\hbox{\bigmsbm C}^{4}/\Gamma$ orbifold singularities}

The essential idea behind the construction of effective field theories 
for $D$--branes (at singularities) is that the fields describing their 
degrees of freedom are related to fundamental strings stretched 
between the branes.
This is indeed how the right gauge group and superpotential is chosen.

This cannot surely happen for the fundamental objects in $M$--theory: 
the $M2$--branes.
The picture just described is no longer valid since there are no 
strings in the eleven--dimensional theory.

The way to overcome this obstacle is to think of $M$--theory as 
 the strong--coupling limit of type $IIA$ string theory.
If this is allowed, then one can try to find the corresponding ten--dimensional 
configuration and see how to describe the low energy effective field 
theory for this latter.
It has been shown \cite{D2D6} that $M2$--branes at orbifold 
singularities arise as a particular phase in the diagram describing a 
more complex situation where one has to deal with $D2$--branes of 
type $IIA$ theory localized onto $D6$--branes, localized or smeared 
$M2$--branes and various other field--theory phases.

In particular, if one studies $N_2$ $D2$--branes over 
$N_6$ $D6$--branes, when $N_6 \ll N_2$, the effective description is 
that of $M2$--branes at an $A_{N_6-1}$ singularity $\IC^2/\IZ_{N_6}$.

The fact that one obtains such a singularity can be understood by the 
supergravity solution representing these
$D2$--branes localized within $D6$--brane in the decoupling limit \cite{local}.
This solution  is given by a Minkowski space $M^{(10,1)}$ (with one direction 
compactified) with $\IZ_{N_{6}}$ identifications over four dimensions.
If we call $\a$ the $N_6$--th root of unity and we complexify 
the four real dimensions on which we make the identification, 
the identification we have to perform is 
\beq
(z_1,z_2) \sim ( \a \, z_1, \a \, z_2).
\eeq
This kind of action means that we have a Gorenstein canonical 
singularity only if $N_6 = 2$ and thus we have to restrict ourselves 
to orbifold singularities of the form 
$$
\frac{\IC^4}{\IZ_2 \times \Gamma^{\prime}}.
$$
According to what we have just said, the $\IZ_2$ action must be 
chosen to be $( -, -, +, + )$ while the $\Gamma^{\prime}$ action must be of 
the form $(\omega^{a_1}, +, \omega^{a_2}, \omega^{a_3})$ because this 
is the only action leading to a Gorenstein canonical singularity, and
compatible with a consistent compactification down 
to ten dimensions. 

Given these restrictions and limiting the analysis to cyclic groups, 
we have found only two classes of Gorenstein 
canonical singularities of the form\footnote{A recent paper \cite{AhnKim}  
studies a singularity of the form $\IC^4/(\IZ_2 \times \IZ_2 \times \IZ_2)$ with  action  
given by $g_1 = (-, -, +, +)$, $g_2 =(-, +, -, +)$ and 
$g_3 =(-, +, +, -)$. But this means that one has to quotient $\IC^4$ 
also by the composite generator $g_1 g_2 g_3 = (-, -, -, -)$ which has 
exactly the form of the quotient $\IC^4/\IZ_2$. This is known to be a 
{\it terminal} singularity and thus admits no Calabi--Yau 
resolutions.} $\IC^4/(\IZ_2 \times 
\Gamma^{\prime})$ satisfying the above requirements.
These are given by
\beq
\frac{\IC^4}{\IZ_2 \times \IZ_{2m}} \qquad \hbox{ and } \qquad 
\frac{\IC^4}{\IZ_2 \times \IZ_{2m} \times \IZ_{2m}}, 
\eeq
with $m \geq 2$.
The first is chosen with an action given by $g_1 = (-,-,+,+)$ and 
$g_2 = ( -, +, \omega^{m-1}, \omega)$; the second with action 
$g_1 = (-,-,+,+)$, $g_2 = ( \omega^{2m-1}, +, 
\omega, +)$ and 
$g_3 = ( \omega^{2m-1}, +, +, \omega)$, with $\omega$ the $2m$--th 
root of the unity.

\bigskip

{$\bullet$ \it The orbifold $\; \dfrac{\IC^4}{\IZ_2 \times 
\IZ_{2m}}$, $m\geq 2$}

\bigskip

The toric data for the $\IC^4/(\IZ_2 \times \IZ_{2m})$ orbifold can be 
deduced by one of the recipes described in \cite{Greene} or (chapter 
2 of) \cite{Fulton} and
are contained in the following matrix:
\beq
\cA = \left( 
\bet{ccccccc}
1 & 1 & 0 & $- m$ & 1 & 0 & 0 \\
0 & -2 & 0 & $2m$ & -1 & 1 & 0 \\
0 & 0 & 1 & $1-m$ & 0 & 0 & 0  \\
0 & 2 & 0 & 0 & 1 & 0 & 1 
\eet\right).
\eeq
From $\cA$ one can derive the charge matrix $Q$ \cite{MorrisonPlesser}, 
which is
\beq
Q = \left( 
\bet{ccccccc|c}
$m$ & 0 & $m-1$ & 1 & 0 & $-2m$ & 0 & $\zeta_1$\\
1 & 0 & 0 & 0 & -1 & -1 & 1 & $\zeta_2$ \\
0 & 1 & 0 & 0 & -1 & 1 & -1 & $\zeta_3$ 
\eet\right),
\eeq
with an extra column including the $D$--terms.

This simple matrix allows us to see that there are no interesting 
resolutions of this singularity (for any $m$)
leading to the  $C(Q^{111})$ or $C(M^{110})$ cones. 
We can find partial resolutions giving some of the  lower dimensional 
singularities described in \cite{MorrisonPlesser}, like the conifold 
($\zeta_2 = 0$ or $\zeta_3 = 0$), the suspended pinch point 
($\zeta_2 = 0$ and  $\zeta_3 = 0$) and the $\IZ_2$ orbifold 
singularities ($\zeta_2 + \zeta_3 = 0$).
All these imply that now $\IC^4/ \Gamma$ is at least reduced to $\IC\times 
\IC^3/\Gamma$.

\bigskip

{$\bullet$ \it The orbifold $\; \dfrac{\IC^4}{\IZ_2 \times \IZ_{2m} \times 
\IZ_{2m}}$, $m\geq 2$}

\bigskip

The toric data for the $\IC^4/(\IZ_2 \times \IZ_{2m} \times \IZ_{2m})$ orbifold 
are contained in the following matrix:
\beq
\cA = \left( 
\bet{ccccccccccc}
1 & -1 & $1 - 2m$ & $1 - 2m$ & 0 & 0 & 0 & -1 & -1 & -1 & -2 \\
1 & 2 & 0 & 0 & 1 & 0 & 0 & 1 & 1 & 0 & 1 \\
1 & 0 & $2m$ & 0 & 0 & 1 & 0 & 1 & 0 & 1 & 1 \\
1 & 0 & 0 & $2m$ & 0 & 0 & 1 & 0 & 1 & 1 & 1 
\eet\right),
\eeq
and we can derive again the charge matrix with the $D$--terms:
\beq
Q = \left( 
\bet{ccccccccccc|c}
1 & 0 & 0 & 0 & -1 & -1 & 0 & 1 & 0 & 0 & 0 & $\zeta_1$\\
1 & 0 & 0 & 0 & -1 & 0 & -1 & 0 & 1 & 0 & 0 & $\zeta_2$\\
1 & 0 & 0 & 0 & 0 & -1 & -1 & 0 & 0 & 1 & 0 & $\zeta_3$\\
1 & 0 & 0 & 0 & -1 & 0 & 0 & 0 & 0 & -1 & 1 & $\zeta_4$\\
0 & 1 & 0 & 0 & -1 & 1 & 0 & -1 & 0 & 0 & 0 & $\zeta_5$\\
$2m - 1$ & 0 & 1 & 0 & 0 & $-2m$ &  0 & 0 & 0 & 0 & 0 & $\zeta_6$\\
$2m - 1$ & 0 & 0 & 1 & 0 & 0 & $-2m$  & 0 & 0 & 0 & 0 & $\zeta_7$
\eet\right).
\eeq

This reveals to be the right choice to obtain the desired conifold 
resolution.
The cone over the $Q^{111}$ manifold \eqn{Amatrix} can indeed be obtained 
by partially resolving 
this singularity\footnote{In \cite{OhTatar} $C(Q^{111})$ was obtained 
from the resolution of the factorised 
${\msbms C}^2/{\msbms Z}_2 \times {\msbms C}^2/{\msbms Z}_2$, 
claiming that $Q^{111}$ is topologically a trivial $S^3$ bundle over 
$S^2 \times S^2$.} keeping $\zeta_2- \zeta_3 = 0$ and $\zeta_4 = 0$.
We recognize in these two rows
$$
\left(
\bet{cccccc|c}
1 & 1 & 1 & 1 & -2 & -2 & $\zeta_2- \zeta_3 = 0$ \\
1 & 1 & -1 & -1 & 0 & 0 & $\zeta_4 = 0$
\eet
\right),
$$
the \eqn{Amatrix} data.

Again, we can find many $\IC\times \IC^3/\Gamma$ singularities such 
as the conifolds, the $\IZ_2$ orbifolds and the suspended pinch points.
The still missing resolution is the one leading to $C(M^{110})$.
It seems that the only way to find such a partial resolution is 
to look at configurations of $M2$--branes at singularities which 
do not admit a ten--dimensional description.

\section{Some comments on the gauge theories}

Now that we have outlined the $D$--terms and resolutions necessary to 
obtain the $C(Q^{111})$ conifold from an orbifold, we would like to derive 
the gauge theory by "resolving" the orbifold gauge theory.
As already said, for $D$--branes in ten--dimensional string theory 
the orbifold theory is derived from the theory on the smooth covering 
space by a suitable projection, and the charge matrices we have used to find 
the resolutions tell us which fields have to acquire a vev.

Unfortunately, in our case, the only information we can can have is about which is the 
correct orbifold to partially solve, but the derivation of the field 
theory has to be found in a more indirect way.

We have to perform a double projection, the first to quotient the 
smooth $\IC^4$ to $\IC^4/\IZ_2$ and the second to derive the 
$\IC^4/(\IZ_2 \times \Gamma^{\prime})$ we are to study.
The first step is accomplished by passing from the eleven--dimensional 
system of $M2$--branes at a $\IC^2/\IZ_2$ singularity to the ten--dimensional 
one, involving $D2$--branes on {\it two} $D6$--branes. 
This yields automatically the projected theory, with no need to 
perform the projection by hand.

Depending on the $\Gamma^{\prime}$ chosen, we will have to study the 
field theory describing the low energy limit of 
$N k$ (where $k = |\Gamma^{\prime}|$) 
$D2$--branes over two $D6$--branes.
We will choose the $D2$--branes to lie in the $x_1$, $x_2$ directions, 
while the $D6$--branes will be stretched in the $x_1, \ldots, x_6$ 
directions.
Once reduced to such a configuration in ten dimensions, 
$\Gamma^{\prime}$ will act on the $x_3, \ldots , x_8$ directions with 
the proper action.

The coordinates transverse to the $D6$--branes are non--dynamical 
degrees of freedom as seen from the $D2$--branes point of view. 
If we call indeed $g_3$ the effective coupling constant for the fields on the 
$D2$ and $g_7$ that for the ones on the $D6$--branes, these have to be 
related by  
$$
g_3^2 = \frac{g_7^2}{V_{3456}},
$$
where $V_{3456}$ is the $D6$ volume transverse to the $D2$--brane.
Now, sending $V \to \infty$, the kinetic energy of these fields explodes 
to infinity and thus they have to be treated as classical degrees of 
freedom frozen at a specific value.

On the $D2$--branes there lives an $\cN = 8$ gauge theory, whose content is 
given by the gauge field $A^{\mu}_{ij}$ ($\mu = 0,1,2$) and the 
transverse coordinates fluctuations 
$\Phi^{I}_{ij}$ ($I = 3,\ldots,9$) sitting in the 
adjoint of $U(Nk)$.
This is exactly the same superconformal field theory which lives on 
an $M2$--brane placed on flat space \cite{M2}. 
There we had a theory with eight scalars, but it is easily seen that 
in three dimensions one can dualize one of these scalars to obtain a 
vector field.
Since these $D2$--branes are placed onto two $D6$--branes, 
supersymmetry is broken down to $\cN = 4$ and the $\Phi^{I}_{ij}$ 
have to be split in those of the new vector multiplets ($A^{\mu}_{ij}, 
\Phi^7_{ij}, \Phi^8_{ij}$ and $\Phi^9_{ij}$) and those of 
the hypermultiplets ($X_{ij}^{34}, X_{ij}^{56}$).

These are not the only fields living on the $D2$--branes.
We also have to consider the strings stretched among the $D2$ and 
$D6$--branes. 
These carry a $U(Nk)$ color and a $U(2)$ flavour index 
and, from the three--dimensional point of view, they can be seen 
as quark fields $Q_{iA}$ and $\widetilde{Q}_{i\dot{A}}$ 
(where $ A \in {\bf 2} \in U(2)_{fl.r}$ and $\dot{A} 
\in \bf\bar{2}$).

We could then write the superpotential for such a theory as it is 
derived from the tree--level stringy interactions.
This theory will then be the superconformal field theory describing 
the low--energy limit of a system of $Nk$ $M2$--branes 
placed at the $\IC^2/\IZ_2$ orbifold singularity.

We chose $Nk$ and not simply $N$ branes, because now we want to 
perform the projection needed to obtain the theory on $\IC^4/(\IZ_2 
\times \Gamma^{\prime})$.

At this point indeed, if we choose $g \in \Gamma^{\prime}$ and denote 
its actions on $\IC^{4}$  and on the Chan--Paton factors by $R(g)$ 
and $S(g)$ respectively, we can determine which are the surviving 
fields for the projected theory.
The surviving components \cite{Greene} 
of the gauge field $A$ will be the ones for which 
\beq
A_{ij}^{\mu} = (S(g)A^{\mu}S^{-1}(g))_{ij}
\eeq
and the surviving scalars must satisfy
\beq
\begin{array}{rcl}
    \Phi^I_{ij} &=& {R^I}_J(g) (S(g) \Phi^J S^{-1}(g))_{ij}, \\
    X^I_{ij} &=& {R^I}_J(g) (S(g) X^J S^{-1}(g))_{ij}.
\end{array}
\eeq

We also have to consider the quark fields $Q$ and $\widetilde{Q}$ 
which are again to be projected under the $\Gamma^{\prime}$ action.
This means that we will keep only the components invariant under
\bea
Q_{iA} &=& (S(g) Q)_{iA}, \nonumber \\
\widetilde{Q}_{i\dot{A}} &=& (\widetilde{Q} S^{-1}(g))_{i\dot{A}}.
\eea

One then has to substitute the surviving fields into the $\cN = 4$ 
super Yang--Mills theory describing the $D2$--$D6$ system.
We already discussed which would be the $D$--flatness conditions for 
the unprojected theory, we thus have to add the $F$--flatness 
conditions which now arise from solving the equations deriving from 
the minimization of the superpotential under its variation over the 
moduli fields $(\Phi, X)$.

All these conditions together allow to derive the complete theory, 
while the geometric informations of section two are not sufficient to 
completely determine the gauge theory.

Here we limit ourselves to the above considerations leaving to a 
future work the exact computation of the SCFT Lagrangian and 
possibly the extension of what just said to $\cN = 1$ or $\cN = 3$ 
cases.

Our hope is that such constructions could bring us to the 
"experimental" test  
of the $AdS$/CFT correspondence through the ($AdS$ masses)/($CFT$ weights) 
relations following the lines of \cite{Last}, 
since the Kaluza Klein spectrum on such 
manifolds can be completely determined and it is actually under computation 
\cite{M111}.

\vskip 1truecm
\paragraph{Acknowledgements.}
\  It is a pleasure for me to thank A. Arsie and B. Greene for many fruitful 
suggestions and, with A. Ceresole, for useful comments 
on a preliminary version of this paper.
I  would also like to thank R. D'Auria and S. Ferrara for helpful 
discussions and B. Sturmfels for mail correspondence.
I would also like to thank all the colleagues at the DFT and in particular
Prof. Fr\'e for his critical reading of the manuscript. 

%
%

\end{document}